%% file: main.tex
\documentstyle[aps,preprint,pre,citesort,epsf,graphics]{revtex}

\newcommand{\dotprod}{\,{\scriptscriptstyle \stackrel{\bullet}{{}}}\,}

\begin{document}

\input{title-and-abstract}

\section{Introduction}
\label{intro}
\input{intro}

\section{Details of the Algorithm}
\label{details}
\input{algorithm-details}

\section{Accuracy and Efficiency of the Algorithm}
\label{results}
\input{results}

\section{Conclusions}
\label{conclusions}
\input{conclusions}

\section*{Acknowledgments}
\input{acknowledgements}

\input{figure-captions}



\input{references}
\newpage

\input{tables}

\newpage

\input{figures}

\end{document}

%% file: title-and-abstract.tex
\title{Efficient Algorithm On a Non-staggered Mesh for
Simulating Rayleigh-B\'enard Convection in a Box}

\author{K.-H. Chiam}

\address{Nonlinear and Statistical Physics, California Institute of
  Technology, Mail Code 114-36, Pasadena, CA 91125-3600}

\author{Ming-Chih Lai}

\address{Department of Applied Mathematics, National Chiao Tung
  University, Hsinchu 300, Taiwan}

\author{Henry
  S.~Greenside\cite{email-correspondence}\cite{CNCS-address}}

\address{Department of Physics, Duke University,
  P.~O.~Box 90305, Durham, NC 27708-0305}

\date{\today}

\maketitle

\begin{abstract}
  An efficient semi-implicit second-order-accurate
  finite-difference method is described for studying
  incompressible Rayleigh-B\'enard convection in a box,
  with sidewalls that are periodic, thermally
  insulated, or thermally conducting.
  Operator-splitting and a projection method reduce the
  algorithm at each time step to the solution of four
  Helmholtz equations and one Poisson equation, and
  these are are solved by fast direct methods.  The
  method is numerically stable even though all field
  values are placed on a single non-staggered mesh
  commensurate with the boundaries.  The efficiency and
  accuracy of the method are characterized for several
  representative convection problems.
\end{abstract}


%% file: intro.tex
Experiments over the last three decades have discovered
many fascinating and poorly understood examples of
pattern formation in large-aspect-ratio
Rayleigh-B\'enard
convection~\cite{Bodenschatez00,Cross93}.  Because of
the prominent role that these experiments play in
understanding sustained nonequilibrium
systems~\cite{Cross93} and because many of the observed
phenomena such as spatiotemporal chaos are difficult to
analyze mathematically~\cite{CrossHohenberg94}, there
is a need to develop computer codes that can simulate
these experiments {\em quantitatively} so that theory
and experiment can be compared with one another. Once
validated, such codes can further be used to explore
regimes not easily attained by experiment such as low
Prandtl number, and to calculate quantities that are
difficult to deduce from experimental data, such as
mean flows~\cite{Lai00anl,Chiam03} and fractal
dimensions~\cite{Egolf00Nature}.

The regime of large aspect ratio~$\Gamma$ (ratio of
horizontal fluid width to fluid depth) poses
significant computational challenges.  Many numerical
degrees of freedom (basis functions or mesh points) are
needed to represent the spatial features of the fluid
and often the dynamics needs to be studied over long
times (many multiples of the horizontal thermal
diffusion time~$\tau_h = \Gamma^2 \tau_v$,
where~$\tau_v=d^2/\kappa$ is the vertical thermal
diffusion time defined in terms of the fluid depth~$d$
and fluid thermal diffusivity~$\kappa$) to insure that
nontransient behavior is being observed. Since the
largest time step allowed by numerical stability for
explicit or semi-implicit algorithms (the ones most
commonly used in Navier-Stokes calculations) is
typically ~$0.05\,\tau_v$ or smaller, simulations in a
representative~$\Gamma=50$ cell may require $10^5$~or
more time steps to eliminate a transient and then study
the statistically stationary properties of the
asymptotic dynamics.  The many degrees of freedom and
long integration times together with the need to repeat
runs for different parameter values imply that
efficient algorithms are essential for studying the
large-aspect-ratio regime.

Because of these computational challenges, there have
been few simulations of three-dimensional
Rayleigh-B\'enard convection with aspect ratios
exceeding~10.  Recent calculations with~$\Gamma$ as
large as~64 have been carried out by Pesch and
collaborators, who used a pseudospectral code on serial
and parallel computers to study spiral defect chaos,
rotating convection, and other
problems~\cite{Decker94SDC,Pesch96,Egolf00Nature}.
However, their code uses periodic boundaries and so can
not take into account quantitatively the influence of
lateral walls on the bulk dynamics.  Arter and
Newell~\cite{Arter88} and Tomita and
Abe~\cite{Tomita99} have carried out simulations in
large boxes with thermally insulated no-slip sidewalls,
the former in a $16 \times 11.5$ aspect ratio box, the
latter in a~$\Gamma=18.84$ square box.  Xi et
al~\cite{Xi97} have studied the transition to
spatiotemporal chaos of a convecting fluid in a
$\Gamma=60$ square cell, but with free-slip horizontal
boundaries that are difficult to achieve
experimentally.  Finally, a Caltech-Duke collaboration
has recently reported results~\cite{Paul01prl,Paul02}
obtained with a parallel spectral element
code~\cite{Tufo99} for aspect ratios up to~30. Their
code can treat quantitatively most geometries and
lateral boundaries used by experimentalists, including
ramps~\cite{Bajaj99}, spoiler fins~\cite{Daviaud89},
and lateral walls of finite thickness and finite
thermal conductivity.  However, the generality of the
spectral element algorithm makes it substantially more
expensive to run than algorithms optimized for a simple
geometry such as a box or cylinder.

In this paper, we introduce and analyze an efficient
semi-implicit finite-difference algorithm for studying
incompressible Rayleigh-B\'enard convection in a box,
with lateral walls that are periodic, thermally
insulated, or thermally conducting.  The code
complements the more flexible spectral element
approach~\cite{Paul01prl,Paul02} by being more than an
order of magnitude more efficient on a serial
processor, for a box with these boundary conditions. It
is well suited for studying long-time dynamics of
small- to moderate-aspect-ratio boxes~\cite{Lai00anl}
($\Gamma \le 20$), with lateral boundaries that are
close to those of many experiments, although not fully
quantitatively accurate since the finite thickness and
finite thermal diffusivity of the lateral walls is not
taken into account.

The main advantages of our algorithm are the simplicity
of implementation and its efficiency on a single
processor.  Its simplicity arises from the use of a
single mesh for all field values.  (This is called a
``non-staggered'' or ``collocated'' mesh in contrast to
a ``staggered'' mesh for which the values of different
fields appear at different points in
space~\cite{Harlow65,Peric88,Armfield91}). A
non-staggered mesh reduces the effort to write and to
validate a code (compared to one using a staggered
mesh), and facilitates porting the code to a
distributed-memory parallel computer.  Earlier work on
Navier-Stokes integrators has suggested that
non-staggered mesh codes can be numerically unstable
because of pressure
oscillations~\cite{Harlow65,Dormy99}. Our results below
show that an algorithm to integrate the Boussinesq
equations on a non-staggered mesh can be numerically
stable.

The use of a single non-staggered mesh also helps to
explain the efficiency of the algorithm.  Using a
standard operator splitting and projection method
together with second-order-accurate finite
differences~\cite{VanKan86,Bell89} on a uniform
three-dimensional mesh, the advancement of the
velocity, temperature, and pressure fields at each time
step requires the numerical solution of four Helmholtz
equations and one Poisson equation.  Because these
elliptic equations and their boundary conditions are
{\em separable}, they can be solved efficiently using
fast direct methods from the FISHPACK
library~\cite{Swarztrauber73,FISHPACK75}, with a
complexity per problem of~$O(N\log(N))$, where~$N$ is
the total number of mesh points. Fast direct methods
are more efficient than most iterative methods on a
single processor~\cite{Botta97}, and have the
additional advantage that no internal parameters need
to be adjusted to obtain convergence.  However, fast
direct methods are not applicable to complex
geometries, to problems with spatially varying
parameters, or to complicated boundary conditions that
lead to nonseparable equations.

The remainder of this paper is organized as follows. In
Section~\ref{details}, we discuss details of our
algorithm, namely how the fields and equations are
discretized and how the resulting equations are solved.
In Section~\ref{results}, we discuss the convergence
properties of the algorithm and its efficiency for
several representative two- and three-dimensional
convection problems. We confirm empirically the
second-order accuracy of the solution and examine how
the largest time step allowed by stability varies with
Prandtl number and with Rayleigh number.  Finally,
Section~\ref{conclusions} presents our conclusions and
suggests some avenues for further algorithmic
improvements. Applications of the algorithm to study
quasiperiodic dynamics and spiral defect chaos in
three-dimensional boxes can be found in
Refs.~\cite{Lai00anl,Chiam03}.


%% file: algorithm-details.tex
\subsection{Equations and Boundary Conditions}
\label{sec:eqs-and-bcs}

Our goal is to integrate the Boussinesq equations that
describe incompressible buoyancy-driven
Rayleigh-B\'enard convection with an external
force~$\bf f$. These equations can be written in the
dimensionless form~\cite{Busse78}
\begin{eqnarray}
  \partial_{t} T(t,x,y,z)
  &=& \left[ - ({\bf v}\dotprod\nabla) T \right]
  \: + \nabla^2 T
  , \label{Eq-Boussinesq-eqs-T}\\
  \partial_{t}{\bf v}(t,x,y,z)
  &=& 
  \left[
  - ({\bf v} \dotprod \nabla){\bf v}
  + \sigma R T \hat{\mathbf{z}}
  + \sigma {\bf f} \right]
  \:+\: \sigma\nabla^{2} {\bf v}
  \: - \nabla p
  , \label{Eq-Boussinesq-eqs-v} \\
  \nabla\dotprod{\bf v}
  &=& 0
  . \label{Eq-Boussiensq-eqs-incompressibility}
\end{eqnarray}
The variables~$x$ and~$y$ denote the horizontal
coordinates, while the~$z$ variable denotes the vertical
coordinate, with the unit vector~$\hat{\bf z}$ pointing
in the direction opposite to the gravitational
acceleration.  The field~${\bf v} = (v_x,v_y,v_z)$ is
the velocity field at point~$(x,y,z)$ at time~$t$,
while~$p$ and~$T$ are the pressure and temperature
fields respectively. The dimensionless
parameters~$\sigma$ and~$R$ denote the Prandtl and
Rayleigh numbers respectively.  The vector field~${\bf
  f}(t,{\bf x},T,{\bf v})$ is some external force,
e.g., a Coriolis force ${\bf f}=2 {\bf
  v}\times\mathbf{\Omega}$ arising from a rigid
rotation of the convection cell with constant angular
velocity~${\mathbf\Omega} = \Omega \hat{\bf z}$.  The
terms grouped in brackets in
Eqs.~(\ref{Eq-Boussinesq-eqs-T})
and~(\ref{Eq-Boussinesq-eqs-v}) are those containing
nonlinear terms or linear terms with low-order spatial
derivatives, and will be integrated explicitly by the
operator splitting method described below.

We would like to integrate
Eqs.~(\ref{Eq-Boussinesq-eqs-T})-(\ref{Eq-Boussiensq-eqs-incompressibility})
in a box geometry defined by the region
\begin{equation}
  \label{Eq-box-geometry}
  -\frac{\Gamma_x}{2} \le x \le \frac{\Gamma_x}{2}, \qquad
  -\frac{\Gamma_y}{2} \le y \le \frac{\Gamma_y}{2} , \qquad
  -{1\over 2} \le z \le {1\over 2} ,
\end{equation}
where~$\Gamma_x$ and~$\Gamma_y$ are the aspect ratios
in the~$x$ and~$y$ directions respectively (the depth
of the fluid has length~1).  A no-slip velocity
condition on all material walls is assumed
\begin{equation}
  \label{Eq-no-slip-bcs}
  {\bf v} = 0 ,
\end{equation}
and the temperature~$T$ is constant on the bottom and
top plates,
\begin{equation}
  \label{Eq-T-horizontal-bcs}
  T = \pm {1 \over 2}  \qquad \mbox{for} \qquad 
  z = \mp {1 \over 2} .
\end{equation}
The code allows the temperature on any opposing pair of
lateral walls to be periodic, or to satisfy on each
lateral wall an arbitrary Dirichlet boundary condition
(e.g., a thermally conducting wall corresponding to a
linear conducting profile of the form~$T=a + bz$,
where~$a$ and~$b$ are constants), or an arbitrary
Neumann condition (e.g., a thermally insulating wall
with~$\partial_n T = 0$, where~$\partial_n$ is the
normal derivative to the boundary at a given point).
To simplify the following discussion, we will consider
only the case of insulating sidewalls
\begin{equation}
  \label{Eq-T-insulating-bc}
  \partial_n T = 0,  \qquad \mbox{on lateral walls,}
\end{equation}
since the other cases involve just simple
modifications.  Although the pressure field~$p$
formally has no associated boundary condition since it
does not satisfy a dynamic equation, we will be
imposing a Neumann boundary condition on~$p$ as we
explain below (see
Eq.~(\ref{Eq-pressure-Neumann-condition})).

\subsection{The Time Integration Method}
\label{sec:time-integration-method}
  
We next discuss the time integration method, since its
structure can be explained before having to specify a
spatial representation for the fields. In the following
subsection, we discuss how the fields and equations are
discretized and the latter solved using
second-order-accurate finite differences on a uniform
spatial mesh.

Our time integration method uses a standard operator
splitting and projection method~\cite{VanKan86,Bell89},
in which the nonlinear terms containing lower-order or
no spatial derivatives are integrated explicitly, then
the linear diffusion operators are integrated
implicitly, and finally the pressure term~$-\nabla{p}$
is integrated to project the velocity field at the next
time step into the space of divergence-free velocity
fields. Operator splitting has two benefits. First, the
evolution equations for~$T$ and for the velocity
components~$v_i$ are decoupled from one another, which
simplifies the overall algorithm and substantially
reduces the total computer memory needed.  Second,
operator splitting allows larger time steps since the
largest time step~$\Delta{t}$ allowed by stability is
bounded by a first power of the spatial
resolution~$\Delta{x}$, rather than by a second power
as would be the case for a fully explicit method.

Let us assume that, at the~$n$th time step $t_n = n
\Delta{t}$ with~$n \ge 0$, initial fields~$ T^n$
and~${\bf v}^n$ are known that are consistent with the
boundary conditions
Eqs.~(\ref{Eq-no-slip-bcs})--(\ref{Eq-T-insulating-bc}).
These fields are then advanced to the future values~$
T^{n+1}$ and~${\bf v}^{n+1}$ at time~$t_{n+1}=t_n +
\Delta{t}$ as follows:
\begin{enumerate}
\item The nonlinear advective term $N_T[T,{\bf
    v}]=-({\bf v}\dotprod\nabla)T$ of
  Eq.~(\ref{Eq-Boussinesq-eqs-T}) is integrated
  explicitly using a second-order-accurate
  Adams-Bashforth method
  \begin{equation}
    \label{Eq-temperature-explicit-step}
    T^\ast = T^n + {\Delta{t} \over 2}
    \left( 3 N_T\left[T^n,{\bf v}^n\right]
      - N_T\left[T^{n-1},{\bf v}^{n-1}\right] \right) ,
    \qquad n \ge 0 ,
  \end{equation}
  to produce an intermediate field~$T^\ast$.
  Here~$T^{n-1}$ and~${\bf v}^{n-1}$ denote field
  values stored from the previous time step~$t_{n-1} =
  t_n - \Delta{t}$. For the first time step~$n=0$ only,
  a second-order-accurate single-step integrator,
  Heun's method~\cite{Kincaid96}, is used in place of
  Adams-Bashforth to avoid the dependence on the
  unavailable field values at time~$t=-\Delta{t}$.
  
\item The intermediate field~$T^\ast$ is then advanced
  to the temperature field~$T^{n+1}$ at time~$t_{n+1}$
  by using~$T^n$ as initial data for an implicit
  Crank-Nicolson step applied to the diffusion term in
  Eq.~(\ref{Eq-Boussinesq-eqs-T}):
  \begin{equation}
    \label{Eq-T-crank-nicolson-eq}
    {T^{n+1} - T^\ast \over \Delta{t} }
    = {1\over 2} \left(
      \nabla^2 T^{n+1} + \nabla^2 T^n
    \right) .
  \end{equation}
  This can be written as a constant-coefficient
  Helmholtz equation for the future
  field~$T^{n+1}$:
  \begin{equation}
    \label{Eq-T-Helmholtz-eq}
    \left( 1 - {\Delta{t}\over 2}\nabla^2
    \right)T^{n+1}
    = T^\ast + {\Delta{t}\over 2}\nabla^2
    T^n ,
  \end{equation}
  and is solved with the boundary conditions
  Eqs.~(\ref{Eq-T-horizontal-bcs})
  and~(\ref{Eq-T-insulating-bc}) applied to~$T^{n+1}$.
  
\item Three similar pairs of explicit and implicit
  steps are then executed successively, first for the
  velocity component~$v_x$, then for~$v_y$, and then
  for~$v_z$.  The explicit steps advance the velocity
  field~${\bf v}^n$ to an intermediate field~${\bf
    v}^{\ast}$, and then the implicit steps
  advance~${\bf v}^\ast$ to a {\em second} intermediate
  field~${\bf v}^{\ast\ast}$. If we denote
  by~$N_i[T,{\bf v}]$ the expressions in brackets of
  Eq.~(\ref{Eq-Boussinesq-eqs-v}) for~$i=x$, $y$,
  and~$z$, then each explicit step has the form
  \begin{equation}
    \label{Eq-vi-explicit-step}
    v_i^\ast = v_i^n + {\Delta{t} \over 2}
    \left( 3 N_i\left[T,{\bf v}^n\right]
      - N_i\left[T,{\bf v}^{n-1}\right] \right) .
  \end{equation}
  Heun's single-step method is again used at
  time~$t_0=0$ to avoid unavailable field values at
  time~$t_{-1}=-\Delta{t}$. Each 
  field~$v_i^n$ is next used as initial data for an
  implicit Crank-Nicolson step that yields a
  constant-coefficient Helmholtz equation for the
  field~$v_i^{\ast\ast}$:
  \begin{equation}
    \label{Eq-vi-Helmholtz-eq}
    \left( 1 - {\sigma \Delta{t} \over 2}\nabla^2
    \right) v_i^{\ast\ast}
    =  v_i^\ast +  {\sigma \Delta{t} \over 2}\nabla^2
    v_i^n .
  \end{equation}
  This is solved with the no-slip boundary condition
  $v_i^{\ast\ast}=0$ on all surfaces,
  Eq.~(\ref{Eq-no-slip-bcs}).
  
\item An incompressible velocity field~${\bf v}^{n+1}$
  at time~$t_{n+1}$ is obtained from the field~${\bf
    v}^{\ast\ast}$ by integrating the final operator
  step
  \begin{equation}
    \label{Eq-operator-pressure-step}
    \partial_t{\bf v} = - \nabla p ,
  \end{equation}
  with initial data~${\bf v}^{\ast\ast}$, followed by a
  projection method~\cite{Chorin67a,Bell89}.  We
  approximate the time derivative in
  Eq.~(\ref{Eq-operator-pressure-step}) with a
  first-order-accurate stencil,
  \begin{equation}
    \label{Eq-discretized-pressure-step}
    { {\bf v}^{n+1} - {\bf v}^{\ast\ast} 
      \over  \Delta{t} }
    = - \frac{1}{2} \left( \nabla p^{n+1} + \nabla p^n \right) ,
  \end{equation}
  apply the divergence operator to both sides, and then
  use Eq.~(\ref{Eq-Boussiensq-eqs-incompressibility})
  in the form $\mbox{$\nabla\dotprod{\bf v}^{n+1}=0$}$.
    This yields a Poisson equation for the pressure
    field~$p$:
  \begin{equation}
    \label{Eq-poisson-eq-for-p}
    \nabla^2 p^{n+1} = - \nabla^2 p^n + {2 \over \Delta{t}}
    \nabla\dotprod{\bf v}^{\ast\ast} .
  \end{equation}
  Once~$p$ is known by solving
  Eq.~(\ref{Eq-poisson-eq-for-p}), we obtain~${\bf
    v}^{n+1}$ from
  Eq.~(\ref{Eq-discretized-pressure-step}) in the form
  \begin{equation}
    \label{Eq-v-n+1-from-p}
    {\bf v}^{n+1} = {\bf v}^{\ast\ast} 
   - \frac{\Delta{t}}{2} \left( \nabla{p}^{n+1} + \nabla{p}^n \right).
  \end{equation}
  
  Although mathematically there is no boundary
  condition for~$p$---and by discretizing first space
  and then time, a boundary condition for~$p$ can be
  avoided as explained in
  Refs.~\cite{VanKan86,Dukowicz92,Perot93}---we will
  solve Eq.~(\ref{Eq-poisson-eq-for-p}) with the
  Neumann condition
  \begin{equation}
    \label{Eq-pressure-Neumann-condition}
    \partial_n p^{n+1} = 0  \qquad \mbox{on all walls,}
  \end{equation}
  since this allows us to use a fast direct method to
  solve Eq.~(\ref{Eq-poisson-eq-for-p}).  There is a
  substantial literature concerning the appropriateness
  and accuracy of the boundary condition
  Eq.~(\ref{Eq-pressure-Neumann-condition})~\cite{Orszag86,Gresho87}.
  Rather than review this literature, we simply point
  out that a Neumann pressure boundary condition has
  been shown by previous researchers to produce
  acceptably accurate results for problems in which the
  fluid is confined by no-slip surfaces, and we show
  directly in Section~\ref{results} that our algorithm
  is second-order accurate in space and second-order
  accurate in time for several representative problems.
\end{enumerate}
The most time consuming part of this algorithm is, by
far, solving the four Helmholtz equations
Eq.~(\ref{Eq-T-Helmholtz-eq}) and
Eq.~(\ref{Eq-vi-Helmholtz-eq}) (for~$i=x$ ,$y$,
and~$z$) and solving the Poisson equation
Eq.~(\ref{Eq-poisson-eq-for-p}).

\subsection{Discretization on a Uniform Mesh}
\label{sec:spatial-discretization}

The explicit and implicit steps of the previous
section---Eqs.~(\ref{Eq-temperature-explicit-step})
and~(\ref{Eq-T-Helmholtz-eq}),
Eqs.~(\ref{Eq-vi-explicit-step})
and~(\ref{Eq-vi-Helmholtz-eq}), and
Eqs.~(\ref{Eq-poisson-eq-for-p})
and~(\ref{Eq-v-n+1-from-p})---are carried out by
discretizing the fields and equations on a {\em single}
non-staggered mesh of points
\begin{equation}
  \label{Eq-xijk-defn}
  {\bf x}_{ijk} = \left(
    i \Delta{x}, j \Delta{y}, k \Delta{z}
  \right) ,
\end{equation}
that is commensurate with the sides of the box
Eq.~(\ref{Eq-box-geometry}).  The mesh
indices~$(i,j,k)$ satisfy
\begin{equation}
  \label{Eq-3d-index-ranges}
  -\frac{N_x}{2} \le i \le \frac{N_x}{2}, \qquad
  -\frac{N_y}{2} \le j \le \frac{N_y}{2}, \qquad
  -{N_z\over 2}  \le k \le {N_z\over 2} .
\end{equation}
The aspect ratios~$(\Gamma_x,\Gamma_y)$ and the
positive integers~$(N_x,N_y,N_z)$ are specified as
input to the code, and the corresponding spatial
resolutions~$(\Delta{x},\Delta{y},\Delta{z})$ are then
determined from the relations~$\Delta{x}=\Gamma_x/N_x$,
$\Delta{y}=\Gamma_y/N_y$, and~$\Delta{z}=1/(N_z)$. For
large-aspect-ratio convection problems, typically
$\Delta{x}=\Delta{y} > \Delta{z}$ since the~$x$ and~$y$
directions are equivalent and there is a finer
structure in the vertical direction caused by the close
opposing horizontal plates.

At all mesh points Eq.~(\ref{Eq-xijk-defn}) interior to
the box, the first- and second-order spatial
derivatives are approximated using centered
second-order-accurate 3-point finite-difference
stencils.  If~$u_{ijk}=u({\bf x}_{ijk})$ denotes the
values of a field~$u({\bf x})$ at the mesh points, then
the partial derivative~$\partial_x u$ at~${\bf
  x}_{ijk}$ is approximated by
\begin{equation}
  \label{Eq-discrete-partial-x}
  \left[ \partial_x u \right]_{ijk}
  \approx { u_{(i+1)jk} - u_{(i-1)jk} \over 2 \Delta{x} } ,
\end{equation}
with similar expressions for~$\partial_y u$
and~$\partial_z u$. The Laplacian of~$u$ at~${\bf
  x}_{ijk}$ is approximated by the usual 7-point
stencil
\begin{eqnarray}
  \label{Eq-discrete-laplacian}
  \left[ \nabla^2 u\right]_{ijk} &\approx &
  { u_{(i+1)jk}  - 2 u_{ijk} + u_{(i-1)jk} \over  \Delta{x}^2 }
  + { u_{i(j+1)k}  - 2 u_{ijk} + u_{i(j-1)k} \over
  \Delta{y}^2 } \\
  & & \qquad
  + { u_{ij(k+1)}  - 2 u_{ijk} + u_{ij(k-1)} \over  \Delta{z}^2 } .
\end{eqnarray}

Nonsymmetric finite-differences are needed to evaluate
expressions on those boundaries for which a Neumann
condition holds (we will call these ``Neumann
boundaries'') since field values outside the domain are
not available. Thus the right side of the Helmholtz
equation, Eq.~(\ref{Eq-T-Helmholtz-eq}), needs to be
evaluated on the Neumann boundaries for which
Eq.~(\ref{Eq-T-insulating-bc}) holds. The value
of~$\nabla^2{T^\ast}$ can be approximated there to
second-order accuracy by using one-sided 4-point
finite-difference approximations for the 2nd-order
derivatives, e.g.,
\begin{equation}
  \label{Eq-4-point-2nd-order-difference}
  [\partial_x^2 T^\ast]_{0jk} \approx
  { 2 T^\ast_{0jk} - 5 T^\ast_{1jk} + 4 T^\ast_{2jk}  - T^\ast_{3jk}
  \over \Delta{x}^2 } ,
\end{equation}
with similar expressions for $\partial_x^2 T^\ast$
at~$x=\Gamma_x$, and for~$\partial_y^2 T^\ast$ on
the~$y=0$ and~$y=\Gamma_y$ boundaries.  The
divergence~$\nabla\dotprod{\bf v}^{\ast\ast}$ on the
right side of the pressure equation,
Eq.~(\ref{Eq-poisson-eq-for-p}), can be approximated to
second-order accuracy using the Dirichlet data
Eq.~(\ref{Eq-no-slip-bcs}) and interior field values by
replacing Eq.~(\ref{Eq-discrete-partial-x}) with the
following 3-point one-sided finite difference
\begin{equation}
  \label{Eq-3-point-1st-order-difference}
  [\partial_x v_x]_{0jk} \approx {-3 (v_x)_{0jk} + 4 (v_x)_{1jk}
  - (v_x)_{2jk} \over 2 \Delta{x} } 
   = { 4 (v_x)_{1jk} - (v_x)_{2jk} \over 2 \Delta{x} }  ,
\end{equation}
with similar expressions for~$\partial_y v_y$
and~$\partial_z v_z$.  The advective derivative~$-({\bf
  v}\dotprod\nabla)T$ in
Eq.~(\ref{Eq-temperature-explicit-step}) vanishes on
these Neumann walls since~${\bf v}$ does, and so the
explicit steps do not require special treatment.

Given the discretizations
Eqs.~(\ref{Eq-discrete-partial-x})-(\ref{Eq-3-point-1st-order-difference}),
the explicit time steps
Eqs.~(\ref{Eq-temperature-explicit-step})
and~(\ref{Eq-vi-explicit-step}) are easily evaluated at
all interior points and on the Neumann boundaries.  For
the implicit steps, the right sides of
Eqs.~(\ref{Eq-T-Helmholtz-eq})
and~(\ref{Eq-vi-Helmholtz-eq}) are also evaluated on
the interior mesh points and on the Neumann boundaries.
These right sides are then used as input to the
FISHPACK~\cite{FISHPACK75} fast direct solvers {\tt
  hw3crt} in three dimensions or {\tt hwscrt} in two
dimensions. Also provided as input to the FISHPACK
solvers are the corresponding boundary conditions,
Eqs.~(\ref{Eq-T-horizontal-bcs})
and~(\ref{Eq-T-insulating-bc}) for~$T$,
Eq.~(\ref{Eq-no-slip-bcs}) for the velocity components,
and Eq.~(\ref{Eq-pressure-Neumann-condition}) for~$p$.
The FISHPACK solvers return second-order-accurate
values (with respect to the spatial resolution) of~$T$,
${\bf v}$, and~$p$ on the mesh~${\bf x}_{ijk}$.

We conclude this section with the observation that the
discrete velocity field~${\bf v}^{n+1}$ obtained from
the concluding step Eq.~(\ref{Eq-v-n+1-from-p}) is only
approximately divergence-free even on the mesh
points~${\bf x}_{ijk}$, i.e., $\nabla\dotprod{\bf
  v}^{n+1} = O(h^2)$ where~$h$ is the larger of the
spatial resolutions~$\Delta{x}$, $\Delta{y}$,
and~$\Delta{z}$.  This is because the discrete
approximation Eq.~(\ref{Eq-discrete-partial-x}) for the
pressure gradient in Eq.~(\ref{Eq-v-n+1-from-p}) is not
consistent with the discretization
Eq.~(\ref{Eq-discrete-laplacian}) used to approximate
the Laplacian~$\nabla \dotprod \nabla$ on the left side
of Eq.~(\ref{Eq-poisson-eq-for-p}). The discrete
Laplacian can be considered as arising from the
evaluation of pressure gradients from pairs of nearest
neighbor points as follows
\begin{eqnarray}
  \label{Eq-Laplacian-in-terms-of-adjacent-gradients}
  [\partial_x^2 p]_{ijk} &=&
  {1 \over \Delta{x}} \left(
    [\partial_x p]_{(i+1/2)jk} - [\partial_x p]_{(i-1/2)jk}
  \right) \\
  &=& {1\over \Delta{x}}\left(
    { p_{(i+1)jk} - p_{ijk} \over \Delta{x} }
    - { p_{ijk} - p_{(i-1)jk} \over \Delta{x} }
    \right) .
\end{eqnarray}
In contrast, Eq.~(\ref{Eq-v-n+1-from-p}) evaluates the
pressure gradient at a point using a finite difference
Eq.~(\ref{Eq-discrete-partial-x}) that spans three mesh
points.


%% file: results.tex
In this section, we discuss several tests that quantify
the accuracy of the above algorithm for a convecting
fluid in a two-dimensional rectangular domain with
periodic sidewalls and in a three-dimensional
rectangular domain with perfectly insulating sidewalls,
Eq.~(\ref{Eq-T-insulating-bc}).  We first confirm the
second-order accuracy of the code with respect to the
spatial and time resolutions by studying how the
temperature and velocity fields converge with
increasing spatial and time resolutions respectively.
We next show empirically how the maximum stable time
step varies with the Rayleigh number~$R$ and the
Prandtl numbers~$\sigma$.  We then calculate the
critical Rayleigh number~$R_c$ and plot the Nusselt
number~$N(R)$ as a function of the Rayleigh number~R,
and obtain good agreement with an analytical
expression~\cite{Schluter65} and with a spectral
code~\cite{Clever74}. Finally, we show the spatial
structure of the fields near onset, to allow comparison
with experiment~\cite{Kirchartz88} and with other
codes.

We note that, on a workstation with a 667~MHz 21264A
64-bit Alpha processor, a square box with aspect ratio
$\Gamma=40$ and spatial resolution
$\Delta{x}=\Delta{y}=\Delta{z} = 1/8$ takes about 4.8~s
per time step of~$\Delta{t}=0.001 \, t_v$. This
corresponds to 80~minutes per vertical diffusion
time~$t_v$ and 90~days per horizontal diffusion
time~$t_h$, so this code is too slow to explore $\Gamma
> 20$ cells over time scales exceeding a horizontal
diffusion time. We discuss two ways of improving the
efficiency of the code in our concluding comments of
Section~\ref{conclusions}.

\subsection{ Second-Order Convergence With Respect to the Spatial
  and Time Resolutions}

We begin by showing that the order of convergence~$p$
of the code is asymptotically~2 (second order) in the
limits of sufficiently fine spatial and time
resolutions. By definition, the convergence with
respect to spatial resolution is of order~$p$ if
$\|u_h-u_{\rm exact}\|=O(h^p)$ in the limit~$h \to 0$,
where~$\|u\|=\sqrt{\sum_{ij} u_{ij}^2}$ denotes the
Euclidean norm of a field~$u$ on the spatial mesh,
$h=\Delta{x}=\Delta{z}$ is the uniform spatial
resolution in the~$x$ and~$z$ directions of a
two-dimensional box, $u_h(x,z)$ denotes a discrete
numerical field on a mesh of resolution~$h$, and
$u_{\rm exact}(x,z)$ is the unknown exact field on the
spatial mesh. By writing $u_h(x,z) = u_{\rm exact}(x,z)
+ C(x,z) h^p$ in the limit~$h \to 0$, for some
function~$C$ independent of~$h$, we deduce that the
order~$p$ can be estimated by examining the quantity
\begin{equation}
  \label{Eq-log2-estimate-of-p}
  p_h =  \log_2\left(
    \|u_{4h} -u_{2h}\|
    \over\|u_{2h} - u_h\| \right) ,
\end{equation}
in the limit~$h \to 0$. The estimate Eq.~(\ref{Eq-log2-estimate-of-p})
involves field values at the three levels of resolution~$4h$, $2h$,
and~$h$, coarsest to finest.  A similar definition for the order of
convergence with respect to time resolution can be made if the spatial
mesh $h$ is replaced by  the time step $g\equiv \Delta t$,
\begin{equation}
  \label{Eq-log2-estimate-of-p-g}
  p_g =  \log_2\left(
    \|u_{4g} -u_{2g}\|
    \over\|u_{2g} - u_g\| \right) ,
\end{equation}

We first studied the convergence with respect to the
spatial mesh~$h$ for a two-dimensional box with
periodic sidewalls, for parameter
values~$\Gamma_x=2\pi/q_c=2.016$, $R=1725.0 \approx
1.01 R_c$, and $\sigma=0.71$. The initial conditions
consisted of a small random perturbation about the
linearly conducting state~$T_0=-z$, ${\bf
  v}_0=(u_0,w_0)={\bf 0}$, and these were integrated
until a stationary state was attained consisting of two
rolls at the critical wave number~$q_c$. For this small
cell, an integration time of $8t_v$ was sufficient for
the dynamics to become stationary.  We then studied the
temperature field, $T_h(x,z)$, and the $x$-component of
the velocity field, $u_h(x,z)$, for different spatial
resolutions $N=\Gamma_x/h=16$, $32$, $64$, and~$128$.
The time step~$\Delta{t}$ was set respectively to the
values $\Delta{t}= 0.01$, $0.005$, $0.0025$,
and~$0.00125$ since the operating splitting makes the
largest stable time step proportional to~$h$.
Table~\ref{Table-L2-convergence-results} summarizes the
values of the limit Eq.~(\ref{Eq-log2-estimate-of-p})
and shows that indeed~$p_h \to 2$ as~$h \to 0$, i.e.,
the code is asymptotically second-order accurate with
respect to the spatial resolution~$h$.

We have also studied the convergence with respect to
the time step~$g$ for a three-dimensional box with
perfectly insulating sidewalls and for parameter
values~$\Gamma_x=\Gamma_y=2$, $R=1725.0 \approx 1.01
R_c$, and $\sigma=0.71$.  The initial condition
consisted of small random thermal perturbations, and
these were integrated up to 20 diffusion times at which
point the state became stationary. For various time
resolutions $g=\Delta t=0.0001$, $0.00005$, and
$0.000025$, all with a space resolution of $N=64$, the
convergence was found (using
Eq.~(\ref{Eq-log2-estimate-of-p-g})) to be $p=1.68$.
This provides evidence that the code is indeed
asymptotically second-order accurate with respect to
the time resolution~$g$.

\subsection{ Dependence of Maximum Stable Time Step on Rayleigh and
Prandtl Numbers}

Since an important practical feature of any production
code is the largest time step that can be taken before
numerical instability occurs, we have studied the
maximum stable time step as a function of the Rayleigh
and Prandtl numbers.  A three-dimensional box with
periodic sidewalls and aspect ratio
$\Gamma_x=\Gamma_y=2$ was used, with a spatial
resolution~$\Delta x = 16$.  The Euclidean norm of the
temperature field, $\|T\|$, was calculated for various
values of~$\Delta t$ each time after a interval of
$20$~vertical diffusion times so that transients
decayed.  The maximum stable time step was then defined
as the value of~$\Delta t$ such that $\|T\|$ remains
bounded, i.e., $\|T\| < 10^5$.

In Fig.~\ref{fig:max_dt}(a), we plot the maximum stable
time step as a function of Rayleigh number for the two
Prandtl number values~$\sigma=1$ (square symbols) and
$\sigma=10$ (cross symbols).  We see that the maximum
stable time step decreases rapidly with increasing
Rayleigh number.  This is to be expected since the
magnitude of the velocity and temperature fields
increase with increasing~$R$.  In fact, a best log-log
fit to the data yields the relation
\begin{equation}
  \max( \Delta t) \propto R^\alpha
\end{equation}
where $\alpha=-1.2$ when $\sigma=1$, and $\alpha=-1.3$ when $\sigma=10$. 

In Fig.~\ref{fig:max_dt}(b), we plot the maximum stable
time step as a function of Prandtl number for fixed
Rayleigh numbers~$R=2048$ (square symbols) and~$R=8192$
(cross symbols).  For~$R=2048$, the maximum stable time
step decreases toward both small and large Prandtl
numbers.  For~$R=8192$, the maximum stable time step
decreases toward small Prandtl numbers but is
approximately constant at large Prandtl numbers.  The
smaller time step needed at small Prandtl numbers can
be attributed to the more dynamical nature of the
convective flow at small Prandtl numbers, such as the
presence of spiral defect chaos~\cite{sdc}.

\subsection{ Estimate of the Critical Rayleigh
Number~$R_c$}

A linear stability analysis of the Boussinesq equations
about the linearly conducting profile between two
infinite horizontal no-slip plates shows that the
critical Rayleigh number $R_c \approx 1707.76$ with
critical wave number $q_c \approx 3.117$, and that the
values of~$R_c$ and~$q_c$ are independent of the
Prandtl number~$\sigma$~\cite{Cross93}.  We tested
these predictions and so validated the code by using a
two-dimensional box of aspect ratio~$\Gamma_x=2\pi/q_c
= 2.016$, with periodic sidewalls, for Prandtl
number~$\sigma=0.71$. We used a uniform spatial
resolution $h=1/N = \Delta{x}=\Delta{z}$ and varied the
number~$N$ of mesh points.

The critical Rayleigh number~$R_c$ was
estimated as the approximate value of~R for which the
growth rate~$\lambda=\lambda(R)$ of a
small-amplitude (0.01) random perturbation about the
linear profile interpolated to zero as a function of~R.
Thus for a sufficiently tiny initial perturbation of
the conducting profile, there is a time interval over
which the $z$-velocity component~$w$ grows
approximately exponentially
\begin{equation}
  \label{Eq-w-growth-rate}
  \|w(t,x,z)\| \approx c(R) e^{\lambda t},
\end{equation}
where $\lambda$ is the growth rate, and~$c$ is
independent of~$t$ but can vary with~R. For~$R >
R_c$, the growth rate is positive, for $R < R_c$,
the growth rate is negative and interpolating between
known positive and negative values provides an estimate
of~$R_c$, for which~$\lambda=0$.

Our protocol was to set~$R=R_+=1725 =
1.01\,R_c$ just above onset, set the
initial velocity field to zero, ${\bf
  v}_0=(u_0,w_0)={\bf 0}$, and set the initial
temperature field~$T_0 = -z + \delta{T}$ to a tiny
random perturbation~$\delta{T}(x,z)$ of the linear
profile~$T=-z$, with $|\delta{T}| \le 0.01$. The
initial conditions were then integrated for a short
time and the growth rate estimated from the formula
\begin{equation}
  \label{Eq-lambda-growth_rate}
  \lambda_+ \approx { \ln\left( \|w(t_2,x,z)\| /\|w(t_1,x,z)\| \right)
    \over t_2 - t_1 } ,
\end{equation}
where~$t_2$ and~$t_1 < t_2$ are two times during the
exponential growth of the magnitude of the
$z$-component of the velocity field~$w$. The
calculation was then repeated with the same initial
condition but for~$R=R_- = 1691 = 0.99 R_c$ to estimate
a decay rate~$\lambda_-$. The critical Rayleigh number
was then estimated as the zero of the line joining the
points~$(R_+,\lambda_+)$ and~$(R_-,\lambda_-)$. The
estimated critical Rayleigh numbers~$R_c$ as a function
of the number of mesh points~$N$ are summarized in
Table~\ref{Table-estimated-Rc}. The values are correct
to a relative error of better than one percent for the
finest spatial resolution, confirming the correctness
and convergence of the discretization and of the
solution technique.

\subsection{The Nusselt number versus Rayleigh
  number Curve~$N(R)$}

Another way to characterize the accuracy of a
convection code is by the dimensionless Nusselt
number~$N(t,R,\sigma)$, which is the instantaneous
global vertical heat transport through the fluid layer,
normalized to the heat transport arising from thermal
conduction alone.  For the dimensionless variables used
in
Eqs.~(\ref{Eq-Boussinesq-eqs-T})-(\ref{Eq-Boussiensq-eqs-incompressibility})
above, the Nusselt number can be expressed in the
form~\cite{Busse78}
\begin{equation}
  \label{Eq-Nusselt-number-formula}
  N = 1 + \langle w \, (T - T_{\rm cond}) \rangle ,
\end{equation}
where~$w$ is the~$z$-component of the velocity field
and~$T_{\rm cond} = -z$ is the temperature profile of
the linear conducting state with~${\bf v}={\bf 0}$. The
brackets~$\langle\cdots\rangle$ denote an average of a
quantity over the horizontal coordinates. Sufficiently
close to onset, numerical values of~$N$ can be compared
with an analytical expression~\cite{Schluter65} that is
valid asymptotically in the limit $R - R_c \to
0^+$.

We have evaluated Eq.~(\ref{Eq-Nusselt-number-formula})
for the two-dimensional domain with periodic sidewalls
of the previous section, with
parameters~$\Gamma_x=2.016$, $\sigma=0.71$, and~$N=16$.
Starting from a small perturbation of the linear
profile, we integrated until a stationary state was
attained, and then evaluated
Eq.~(\ref{Eq-Nusselt-number-formula}) for the
stationary state.
Fig.~\ref{fig:N-vs-R-theory-vs-numer} shows how~$N$
empirically varies with~R, and we compare this curve
with the analytical result of Schl\"uter et
al~\cite{Schluter65}, and with numerical values
obtained by Clever and Busse\cite{Clever74}, who used a
two-dimensional spectral code with periodic side walls.
The agreement is good in both cases and confirms the
correctness and accuracy of the code.

\subsection{Spatial Structure of the Numerical Solutions}

We conclude this section with a few examples of the
spatial structure of the fields obtained from the code,
to show that the fields are physically reasonable when
adequately resolved, and are qualitatively in agreement
with other codes and with
experiment~\cite{Kirchartz88}.

for~$\sigma=0.71$, Fig.~\ref{fig:temperature-contours}
shows contours of constant temperature~$T$ and for two
values of the Rayleigh number, $R=2500$ in~(a) and
$R=10^4$ in~(b).  Warm fluid ascends in the middle of
the cell and descends as cooler fluid on both sides.
As~$R$ increases, a thermal boundary layer forms at the
top and bottom plates, creating a finer spatial
structure that will require eventually a decrease in
the vertical spatial mesh size~$\Delta{z}$.
Fig.~\ref{fig:velocity-field}(a) shows the
corresponding velocity field~${\bf v}=(u,w)$, while
Fig.~\ref{fig:velocity-field}(b) shows the vertical
component~$w$ through the midline of the cell.  The
occurrence of two square-shaped convection cells of
opposite vorticity is in good agreement with
experiment~\cite{Kirchartz88}.

Fig.~\ref{fig:3d-temp-contours} shows constant
temperature contours in a three-dimensional box with
insulating sidewalls at time $t=200 \, t_v$, for
parameters~$\Gamma=16$, $R=2500$, $\sigma=0.71$, and
$h=\Delta{x}=\Delta{y}=\Delta{z}=1/8$,
$\Delta{t}=0.01$. In agreement with
experiment~\cite{Gollub82jfm} and with calculations on
the Swift-Hohenberg model of
convection~\cite{Greenside84pra}, the rolls are
approximately normal to the lateral walls and the
pattern consists of two diagonally opposite foci. For
slightly higher~$R=8500$,
Fig.~\ref{fig:3d-temp-contours}(b) shows that the
oscillatory instability commenced in the form of
ripples that propagate along the length of the rolls.
The occurrence of the oscillatory instability and its
spatial form are in good agreement with the linear
stability analysis of Busse and
collaborators~\cite{Busse78} and with experiment.


%% file: conclusions.tex

We have described and characterized a semi-implicit
finite-difference algorithm for integrating the
Boussinesq equations in two- and three-dimensional
boxes, with sidewalls that are periodic, thermally
insulated, or thermally conducting.  Our approach is
useful for simple geometries like a box, cylinder,
torus, and annulus, with boundary conditions such that
various linear operators are separable so that fast
direct methods can be applied.  The resulting algorithm
is sufficiently efficient that aspect ratios up
to~$\Gamma \approx 20$ can be studied on a
single-processor workstation over several days. We
verified that the code was second-order-accurate with
respect to the spatial and time resolutions, and that
it gave good agreement for the critical Rayleigh number
and for the Nusselt number versus Rayleigh number curve
near onset.

The most significant feature of our algorithm is the
use of a single non-staggered mesh for discretizing the
equations and fields (velocity, temperature, and
pressure).  The use of a single mesh simplifies the
writing and validation of the code, and facilitates
adding new physical terms like a Coriolis force. The
single mesh also allowed the use of fast direct methods
from the FISHPACK library~\cite{FISHPACK75} to solve
the Helmholtz and Poisson equations associated with the
implicit part of each time step. We found that
numerical integrations of the Boussinesq equations were
stable on a single mesh despite results of some
previous papers that suggested that a non-staggered
Navier-Stokes code could be unstable because of
pressure oscillations,

Although the algorithm is useful and has been
successfully applied to several
problems~\cite{Lai00anl,Chiam03}, there are two ways
that the algorithm could be improved for the future
study of large-aspect-ratio Rayleigh-B\'enard
convection. First is to parallelize the code for a
distributed memory parallel computer so that aspect
ratios comparable to the largest experiments ($50 <
\Gamma < 100$) could be studied. This is technically
straightforward and would involve, first, distributing
the arrays that represent the fields over the various
processors, and, second, replacing the fast direct
solvers with iterative methods for sparse matrices.
Because of the simpler data structures and reduced
communication overhead associated with a
finite-difference discretization, the parallelized
algorithm will likely be more efficient for simple
geometries and for simple boundary conditions than the
parallel spectral element method of
Refs.~\cite{Tufo99,Paul02}.

Because time integration algorithms involve sequential
steps, parallelizing a code allows a larger spatial
domain, but not a longer observation time, to be
studied for a fixed amount of wall-clock time. A second
helpful improvement would be to increase the efficiency
of the time integration method close to the onset of
convection so that larger time steps can be taken for a
given computational effort. A weakness of the operating
splitting used in most convection codes---finite
difference, spectral, and spectral-element---is that
the explicit integration of the advection terms imposes
a bound on the time step of the form~$C
\epsilon^{-1/2}\Delta{x}$ where~$C$ is a constant
and~$\epsilon = (R - R_c)/R_c$ is the reduced Rayleigh
number.  This bound is independent of the spatial
resolution and diverges less rapidly in the
limit~$\epsilon \to 0^+$ than the physical time scale,
which is proportional to~$\epsilon^{-1}$ .  It would be
interesting to explore whether a more sophisticated
explicit time-stepping technique such as a matrix
exponential method~\cite{Friesner89,Hochbruck98} or a
fully implicit method~\cite{Liu93jcp,Cross01} may
succeed in allowing larger time steps that are
commensurate with the physical time scale while
retaining the efficiency of the present code.


%% file: acknowledgements.tex
We would like to thank Michael Cross, Paul Fischer, and
Mark Paul for helpful discussions, and the Department
of Energy for supporting this research under grant
DE-FT02-98ER14892.

%% file: figure-captions.tex
\begin{figure}[htb]   
  \caption{
    {\bf (a)} Plot of the maximum stable time step as a
    function of Rayleigh number.  The Prandtl number is
    kept constant at $\sigma=1$ (square symbols) and
    $\sigma=10$ (for cross symbols).  The cell has
    aspect ratio $\Gamma_x=\Gamma_y=2$ and periodic
    sidewalls.  The mesh resolution is $\Delta x=1/16$.
    Small random perturbations in the temperature field
    are used as initial conditions.  The simulation is
    run until $20$~vertical diffusion times, at which
    time the Euclidean norm $\|T\|$ is then calculated.
    The value of $\Delta t$ such that this norm becomes
    greater than $10^5$ is defined as the maximum
    stable time step.  {\bf (b)} Plot of the maximum
    stable time step as a function of Prandtl number.
    The Rayleigh number was kept constant at $R=2048$
    (square symbols) and $R=8192$ (cross symbols).  The
    same aspect ratio, mesh resolution, and initial
    conditions as in (a) were used. }
  \label{fig:max_dt}
\end{figure}

\begin{figure}[htb]   
  \caption{
    {\bf (a)} Comparison of numerical (square symbols) and theoretical
    (cross symbols) Nusselt numbers versus Rayleigh number. The
    numerical values come from time-independent two-dimensional
    nonlinear states (at time~$t = 12$) with periodic sidewalls. The
    parameters have the values $\Gamma_x =2$, $R =2500$, $\sigma =
    0.71$, $\triangle{x}=\triangle{z} = 1/16$ ($N=16$), and
    $\triangle{t} = 0.01$. The initial state was a small random
    perturbation of the linear conducting profile.  The theoretical
    values come from an asymptotic
    expansion~\protect\cite{Schluter65}. {\bf (b)} Comparison of
    Nusselt number versus Rayleigh number obtained numerically for our
    algorithm (squares) and for a spectral code of Clever and
    Busse~\protect\cite{Clever74} ($\ast$).  The agreement is better
    than 3\%.  }
  \label{fig:N-vs-R-theory-vs-numer}
\end{figure}

\begin{figure}[htb]  
  \caption{ {\bf (a)} Contour lines of the temperature
    field~$T(x,z)$ observed at time $t=12$ in a two-dimensional box of
    aspect ratio $\Gamma_x=2$ with periodic sidewalls.  The parameters
    have values $R =2500$, $\sigma = 0.71$, $\triangle{x}=\triangle{z}
    = 1/16$ ($N=16$), and $\triangle{t} = 0.01$.  {\bf (b)} Contour
    lines of the temperature field observed in a simulation using the
    same geometry and spatial resolution as in (a) but for $R=10^4$
    and time step $\triangle{t}=0.0025$. }
  \label{fig:temperature-contours}
\end{figure}

\begin{figure}[htb]  
  \caption{ {\bf (a)} Time-independent velocity field ${\bf
      v}(x,z)$ at time $t=12$ for $R= 2500$, for
    the same geometry and resolution as
    Fig.~\ref{fig:temperature-contours}. The steady
    state consists of two convection rolls at the
    critical wavenumber~$q_c=3.117$. {\bf (b)} Vertical
    velocity component~$w(x,z=0)$ through the midline
    of the cell, indicating the range of~$w$.  }
  \label{fig:velocity-field}
\end{figure}

\begin{figure}[htb]  
  \caption{ {\bf (a)} Weakly time-dependent temperature
    contours at the midplane, $T(x,y,z=0)$, at time~$t=200$ obtained
    from a three-dimensional box of aspect ratio
    $\Gamma_x=\Gamma_y=16$ with no-slip and insulating boundary
    conditions,
    Eqs.~(\ref{Eq-no-slip-bcs})~and~(\ref{Eq-T-insulating-bc}). Parameter
    values are $R=2500$, $\sigma=0.71$, $\Delta{x} = \Delta{y} =
    \Delta{z} = 1/8$, and $\Delta{t}=0.01$.  {\bf (b):} Time-dependent
    temperature contours at the midplane, $T(x,y,z=0)$, for the same
    geometry and resolutions as in (a) but for $R=8500$ for which the
    rolls are unstable to the oscillatory instability, which shows up
    as propagating ripples along the rolls.
    The time-averaged Nusselt number $\langle N
    \rangle = 2.27$ is larger than for {\bf (a)}, for
    which the value is $\langle N \rangle =
    1.44$. }
  \label{fig:3d-temp-contours}
\end{figure}


%% file: tables.tex
\begin{table}
 \caption{
   Estimated order of convergence~$p_h$ from
   Eq.~(\ref{Eq-log2-estimate-of-p}), as a function of
   the number of mesh points~$N=\Gamma_x/h$, for a
   stationary solution of a two-dimensional square box
   with periodic sidewalls. The aspect
   ratio~$\Gamma_x=2.016$, Rayleigh number~$R=1725$,
   and Prandtl number~$\sigma=0.71$. Results are
   presented for for the temperature field~$T(x,z)$ and
   for the $z$-component of the velocity~$u(x,z)$. }
 \label{Table-L2-convergence-results}
 \begin{tabular}{c|cc}
   $N$ & $p_h$~for $T$ & $p_h$ for~$w$ \\
   \hline
   16  & 1.46 & 1.43 \\
   32  & 1.80 & 1.79 
 \end{tabular}
\end{table}


\begin{table}
 \caption{
   Estimated critical Rayleigh number~${\rm
     R}_{\rm c}$, based on where the growth rate~$\sigma=\sigma(R)$
   linearly interpolates through zero. The relative
   error is defined by $({\rm R}_{\rm c} - 1708)/1708$.}
 \label{Table-estimated-Rc}
  \begin{tabular}{c|cc}
   N &  ${\rm R}_{\rm c}$ & Relative error\\ 
   \hline
   16 & 1693.0 & 0.9 \\
   32 & 1696.6 & 0.7 \\
   64 & 1698.5 & 0.5
 \end{tabular}
\end{table}


%% file: figures.tex
\centerline{\epsfxsize=5in \epsfbox{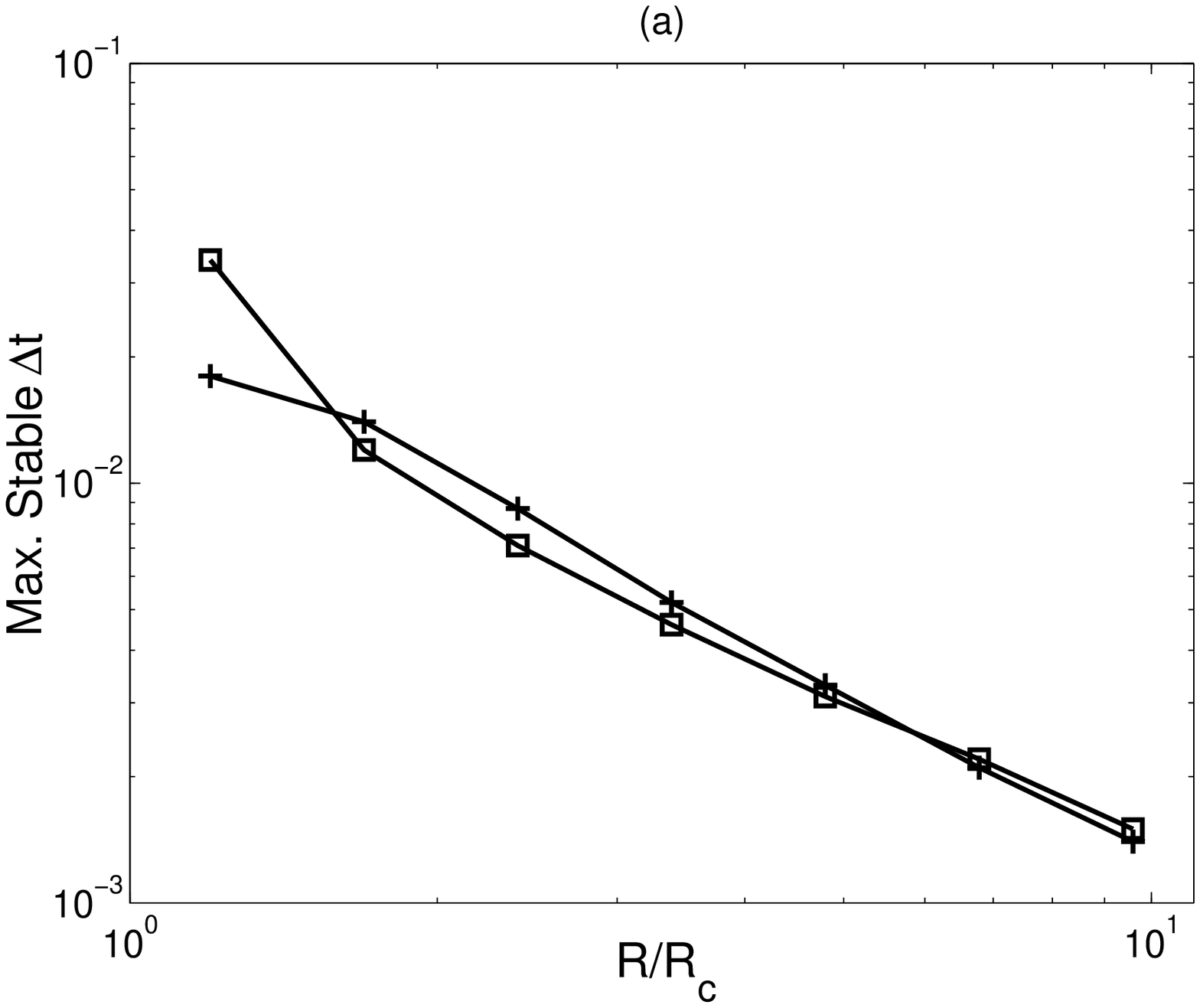}}
\centerline{\epsfxsize=5in \epsfbox{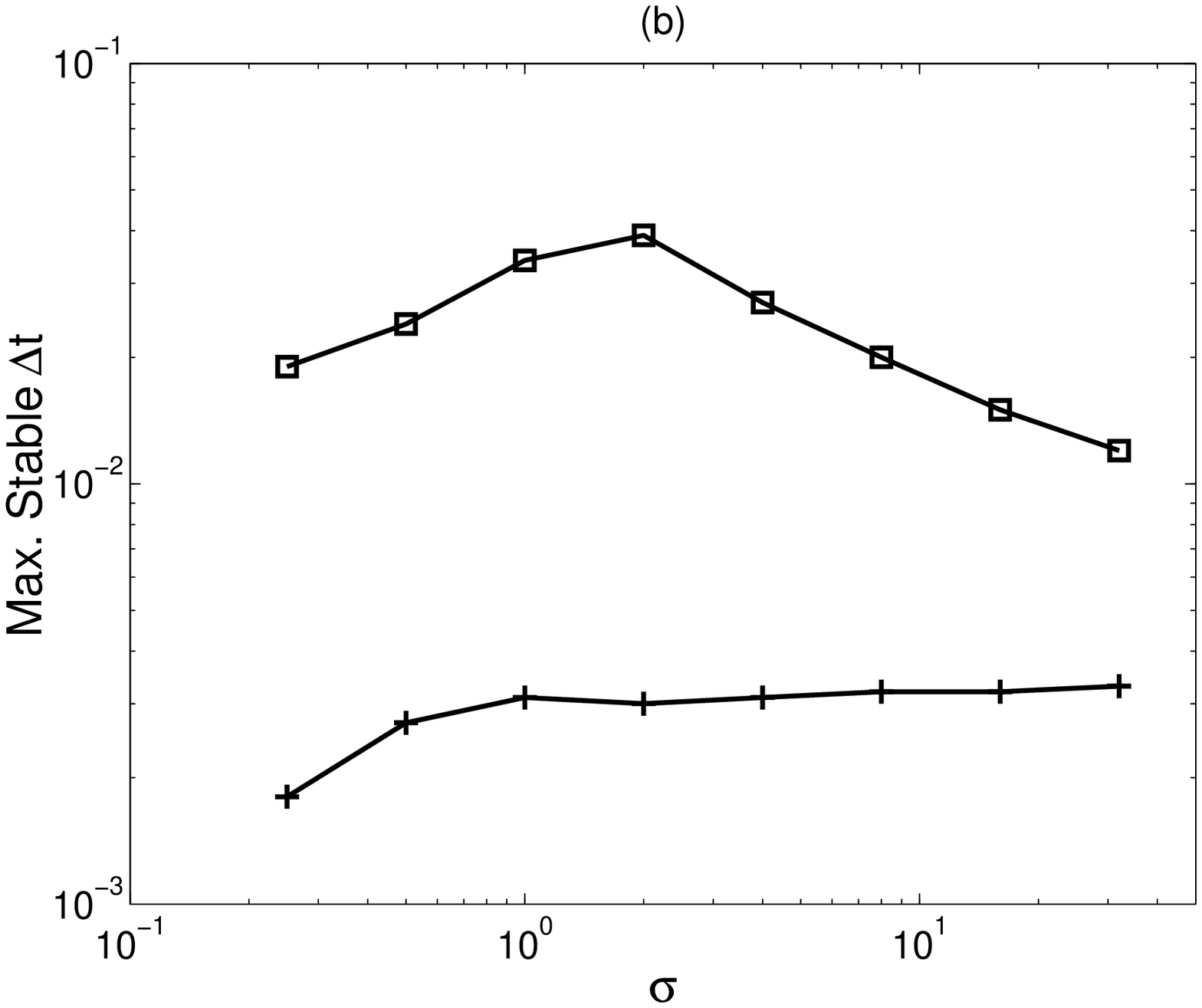}}
\begin{center}
  Figure 1.
\end{center}

\newpage

\centerline{\epsfxsize=5in \epsfbox{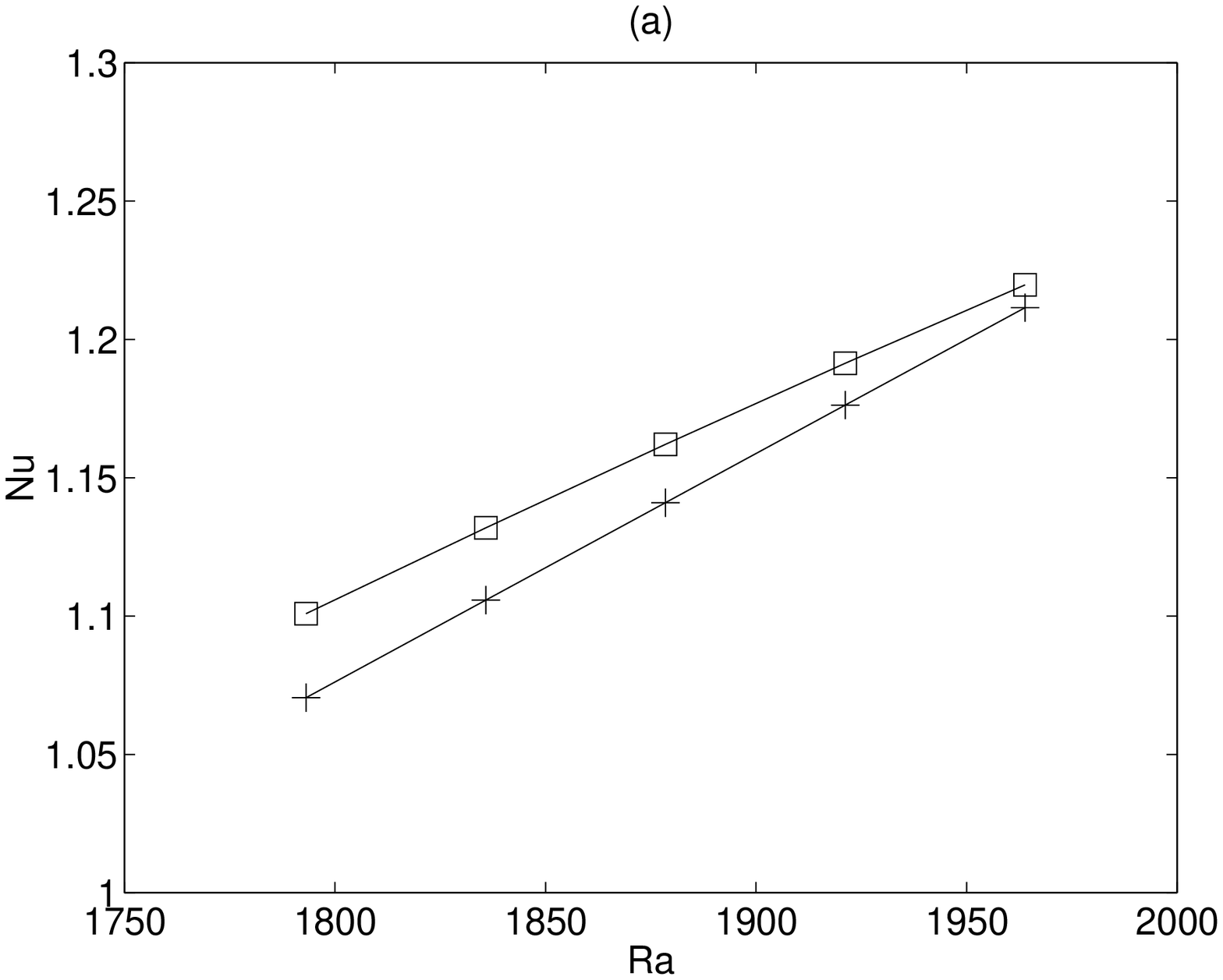}}
\vspace{.3in}
\centerline{\epsfxsize=5in \epsfbox{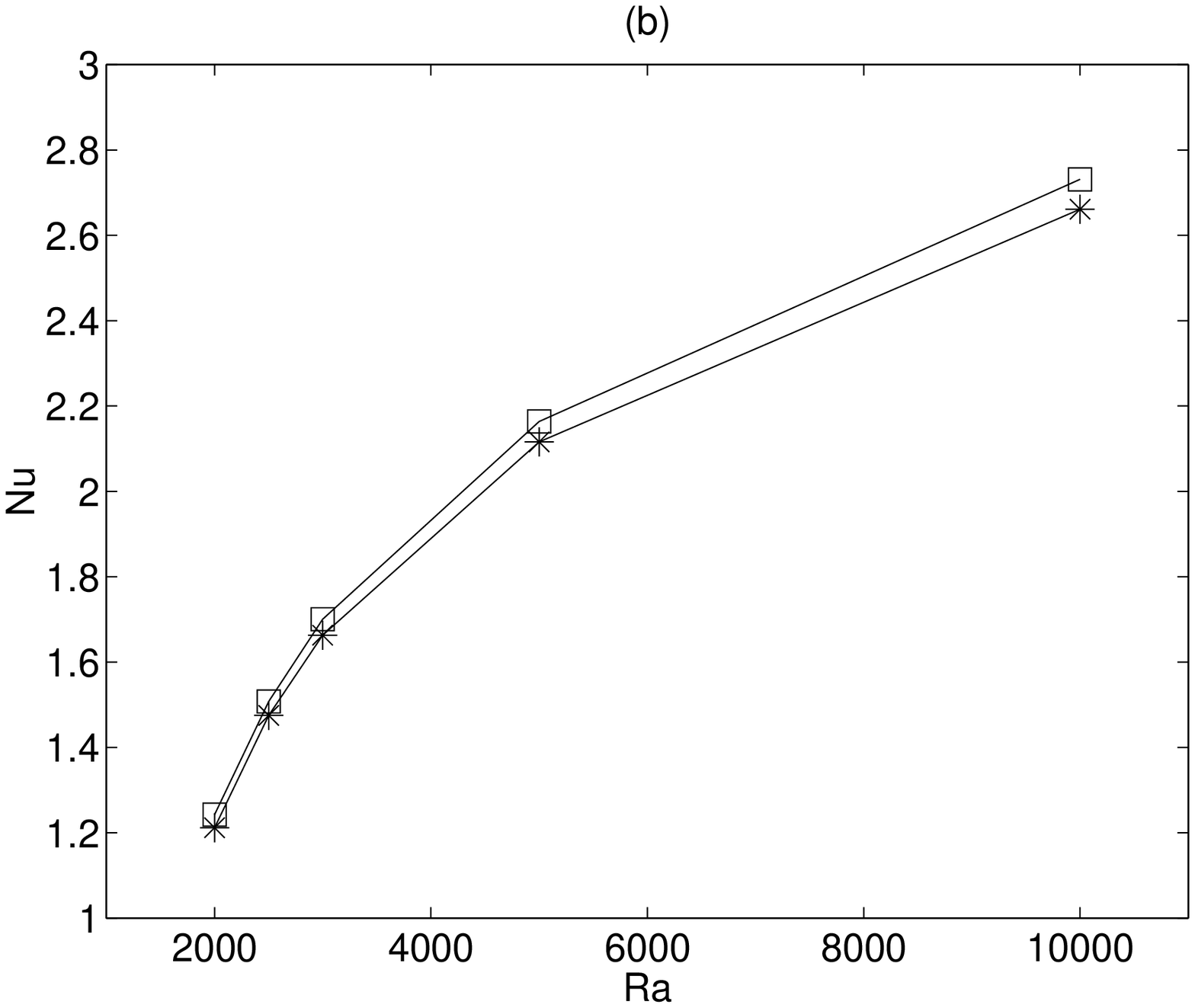}}
\begin{center}
  Figure 2.
\end{center}

\newpage

\centerline{\epsfxsize=6in \epsfbox{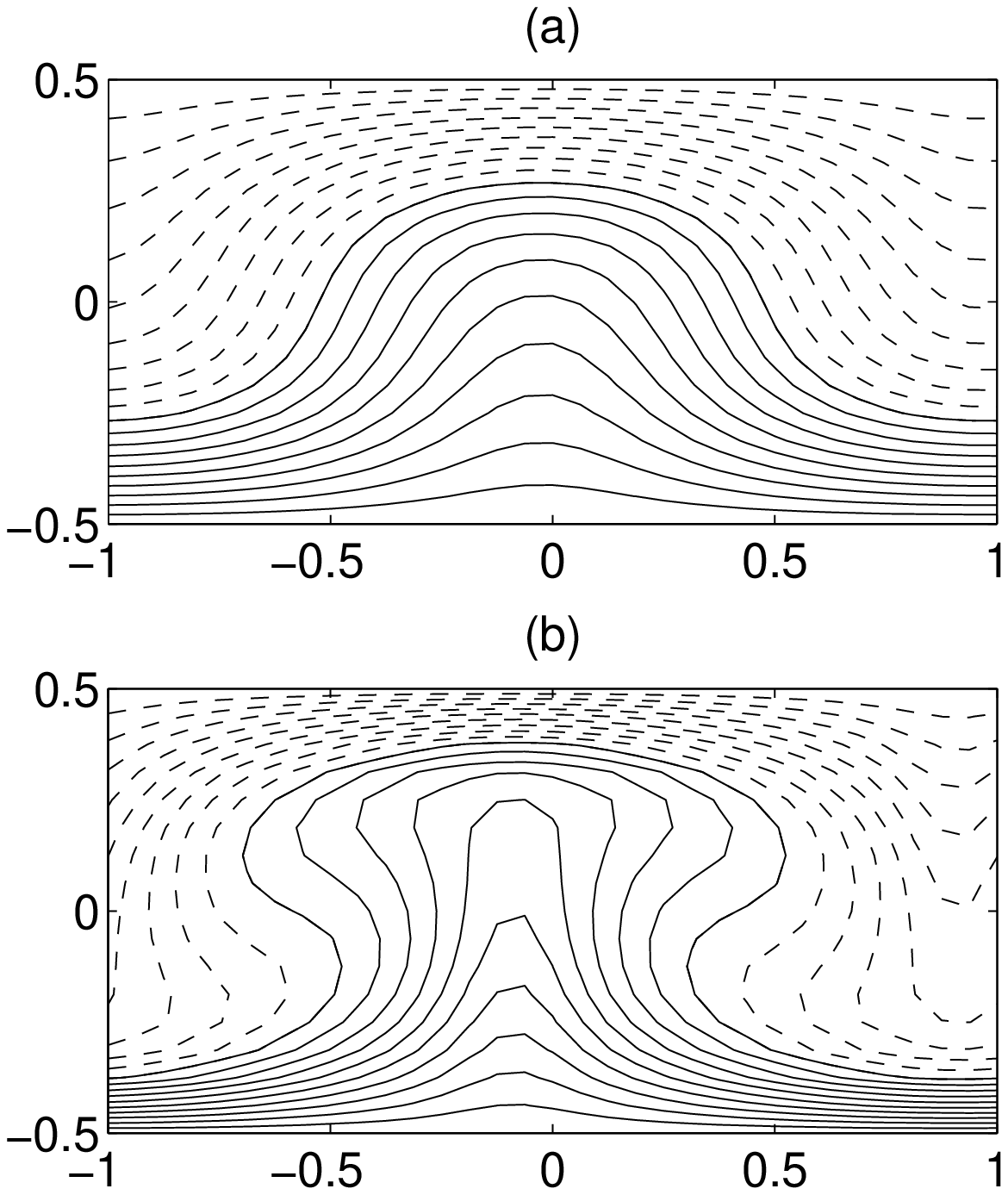}}
\begin{center}
  Figure 3.
\end{center}

\newpage

\centerline{\epsfxsize=5in \epsfbox{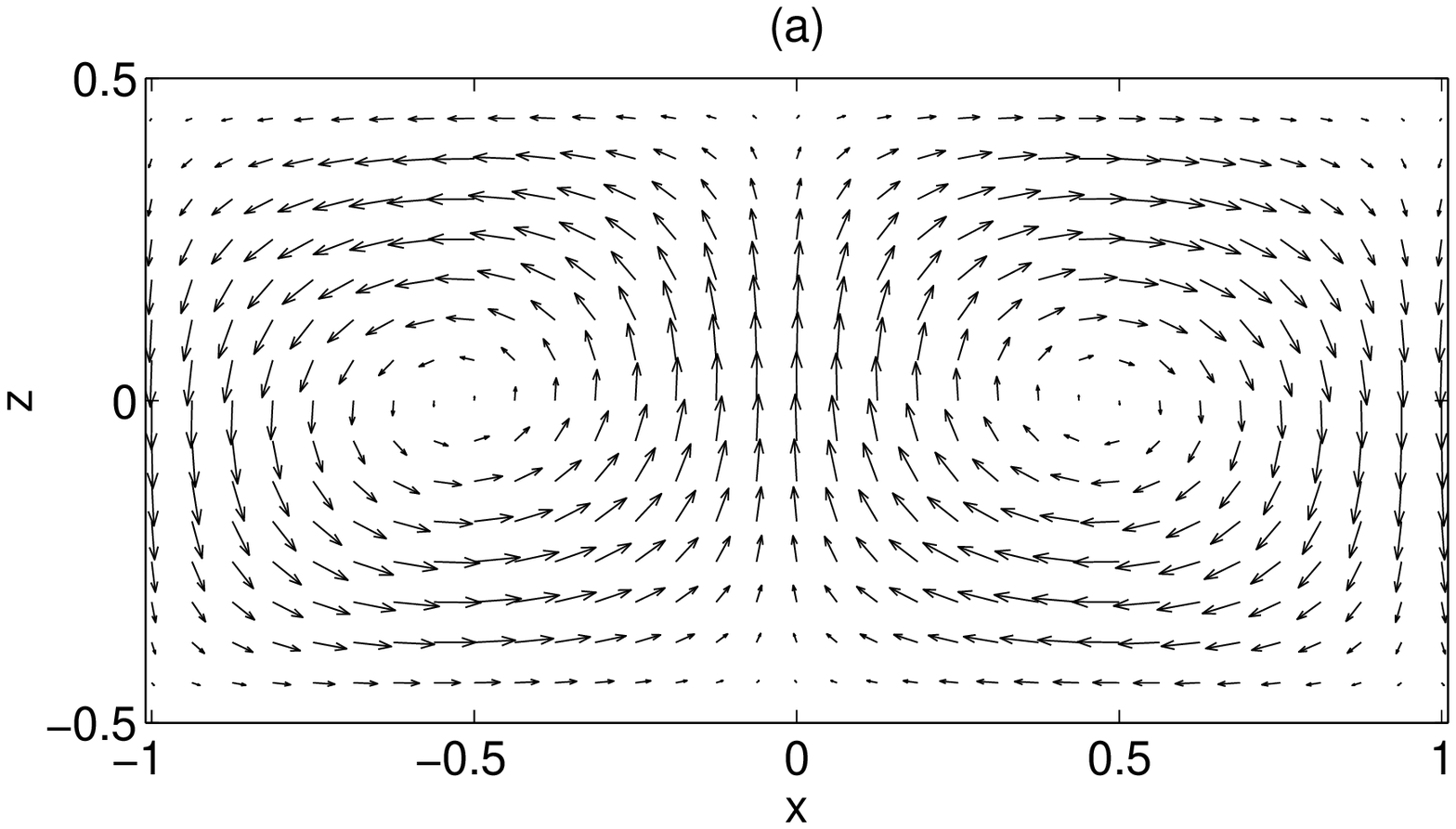}}
\vspace{.3in}
\centerline{\epsfxsize=5in \epsfbox{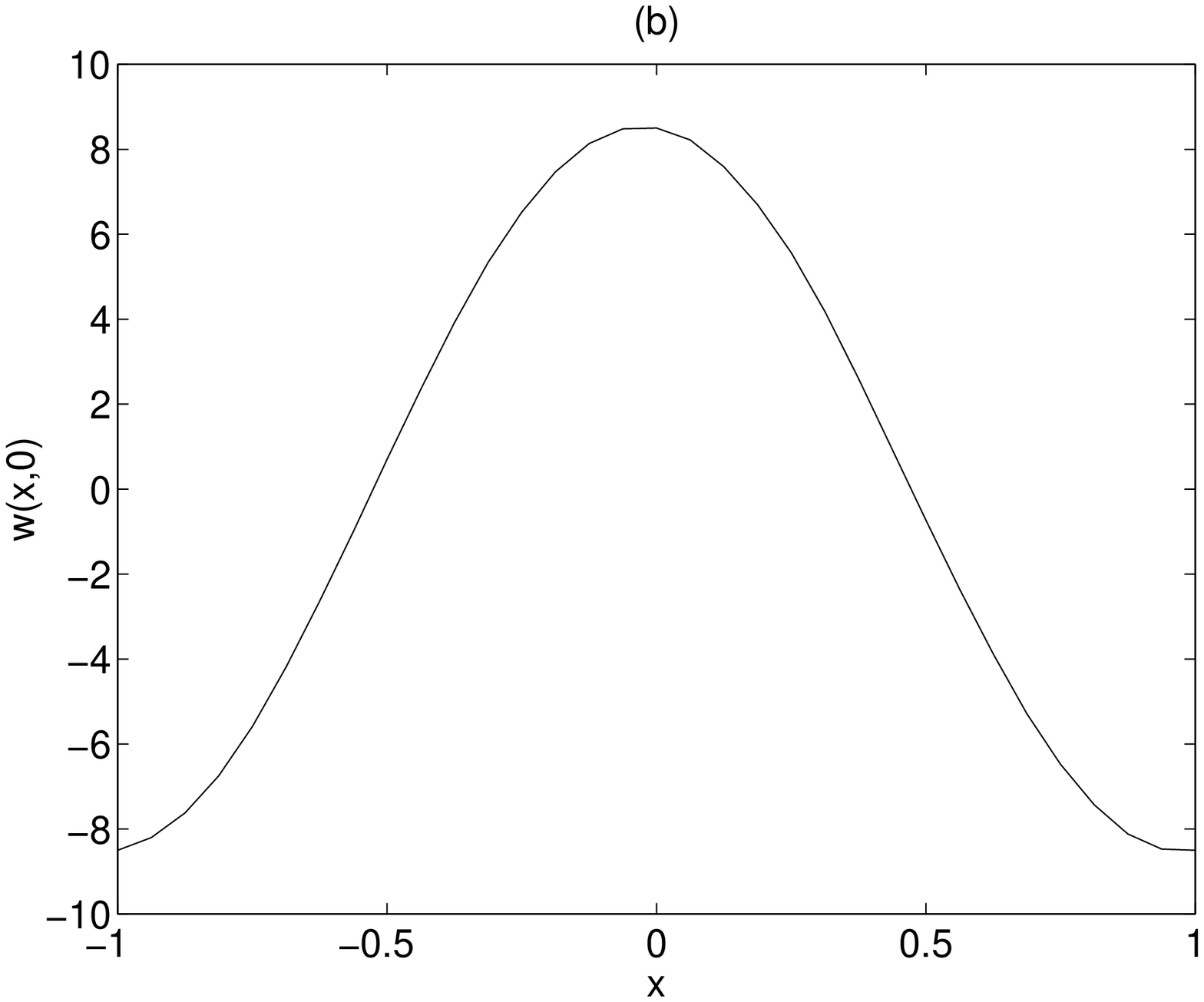}}
\begin{center}
  Figures 4.
\end{center}

\newpage

\centerline{\epsfxsize=4in \epsfbox{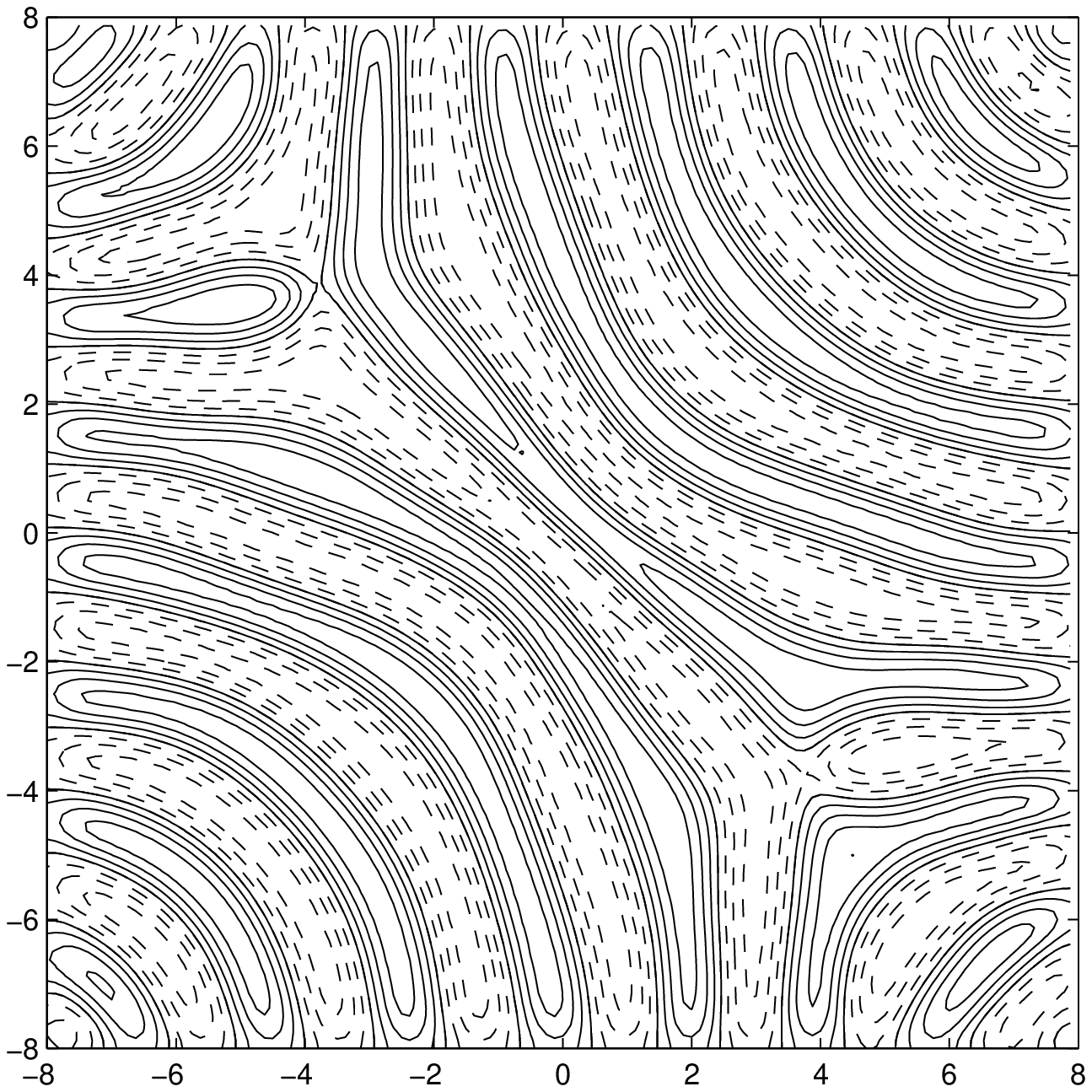}}
\vspace{.3in}
\centerline{\epsfxsize=4in \epsfbox{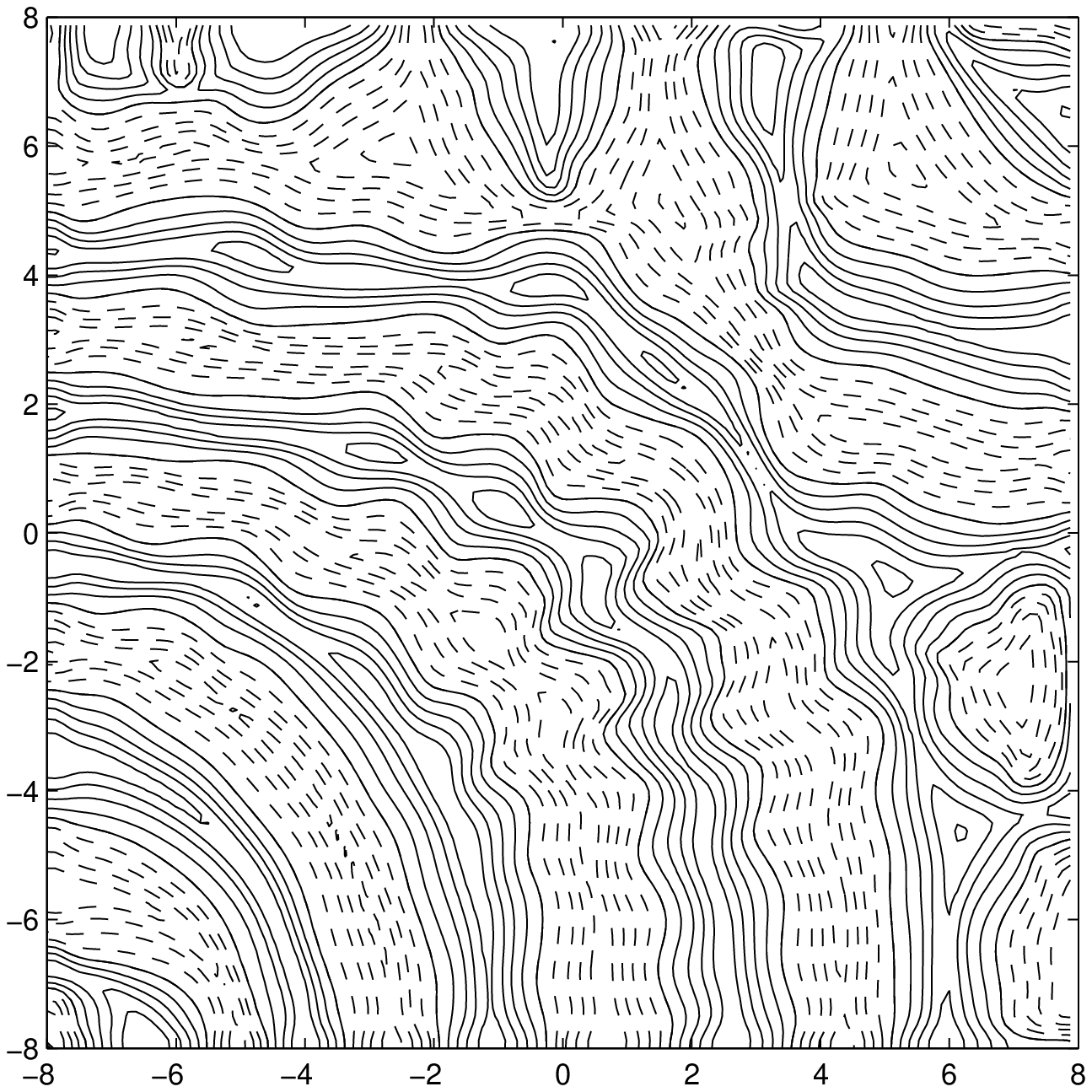}}
\begin{center}
  Figure 5.
\end{center}
